\documentstyle[psfig,12pt]{article}
\def\spacing#1{\renewcommand{\baselinestretch}{#1}\large\normalsize}
\input{epsf}

\def\Item{\par\hang\textindent}
\begin{document}
\pagestyle{myheadings}
\thispagestyle{empty}
\bibliographystyle{unsrt}
\pagenumbering{arabic}
\setcounter{page}{1}
\vspace*{1truein}
\def\@magscale#1{ scaled \magstep #1}
\def\umcp{
        \put(0,16){\line(1,0){4}} \put(5,16){\line(1,0){4}}
        \put(15,16){\line(1,0){4}} \put(10,0){\line(1,0){4}}
        \put(15,0){\line(1,0){4}} \put(20,0){\line(1,0){4}}
        \put(4,0){\line(0,1){16}} \put(5,0){\line(0,1){16}}
        \put(14,0){\line(0,1){16}} \put(15,0){\line(0,1){16}}
        \put(19,0){\line(0,1){16}} \put(20,0){\line(0,1){16}}
        \put(4,16){\oval(8,32)[bl]} \put(5,16){\oval(8,32)[br]}
        \put(14,0){\oval(8,32)[tl]} \put(20,0){\oval(8,32)[tr]}
        \end{picture}}
\def\Item{\par\hang\textindent}

\setlength{\unitlength}{1mm}
\begin{picture}(24,16)(-60,0) \umcp
\vspace{15mm}
\begin{center}
{\Large{\bf Measuring the dynamics of neural
responses in primary auditory cortex}\\[1in]}
Didier A.Depireux, Jonathan Z. Simon and Shihab A.Shamma\\[.1in]
{\em Electrical Engineering Department \& Institute for Systems Research\\
University of Maryland\\
College Park MD {20742--3311}, USA\\
(301) 405-6842\\[1in]}

\end{center}

\begin{center}
We review recent developments in the measurement of the dynamics of the
response properties of auditory cortical neurons to broadband sounds,
which is closely related to the perception of timbre. The emphasis is on
a method that characterizes the spectro-temporal properties of single
neurons to dynamic, broadband sounds, akin to the drifting gratings
used in vision. The method treats the spectral and temporal aspects of
the response on an equal footing.
\end{center}


\tableofcontents
\listoffigures

\section{Introduction}

\subsection{Timbre}

We classify everyday natural sounds by their loudness (related to the
intensity of the sound), their pitch (the perceived tonal height) and
their timbre (the quality of the sound; that which is neither loudness
nor pitch). The perception of timbre, which will be the main focus of
this paper, is what allows us to tell the difference between two vowels
spoken with the same pitch, or the difference between a clarinet and an
oboe playing the same note. When hearing several musical instruments
simultaneously, we can usually tell which instruments are playing by
identifying the different timbres present in the mixed sound. Additionally,
the perception of timbre is quite robust in the presence of noise and
echoes (or reverberations), or even severe degradation such as during a
telephone conversation, in which the sound is severely band-passed. Timbre
perception is therefore an essential attribute of our sense of hearing.

To understand how we extract these different aspects of a sound, we must
unravel what the auditory representation is along the neural pathway. The
approach presented here takes the point of view that the principles used
by neural systems are universal, once the stimulus has reached beyond
the sensory epithelium (whether the cochlea's basilar membrane or the
retina). In particular the ideas presented here are frequently guided
by considering the basilar membrane as a spatial axis, analogous to a
one-dimensional retina, and then using the methods of visual gratings
(drifting and otherwise), to study and characterize cells in the auditory
cortex.

\subsection{Auditory Cortex}

A few general organizational features have long been recognized
in Primary Auditory Cortex (AI), the location of which is shown in
Figure~\ref{fig01}. First is a spatially ordered tonotopic axis, along
which cell responses are tuned from low to high frequencies${}^1$; this is
alternatively called a cochleotopic axis, which reflects the activity along
the cochlea. Note that there are many fields in the auditory cortical
area (the Anterior Auditory Field is shown in Figure~\ref{fig01}), most
of which display a tonotopic organization.

\begin{figure}
\centerline{
\epsfbox{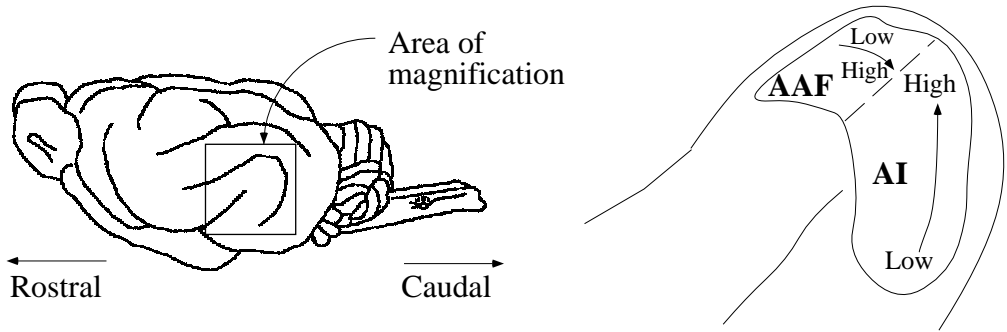}}
\caption[Location and tonotopy of Primary Auditory Cortex]
{\small\spacing{1}\sf The position of the Primary Auditory Cortex (AI) in
the ferret brain. The location of the Anterior Auditory Field (AAF) for
illustration purposes. On the right the tonotopic axis is overlaid for
both AI and AAF.}
\label{fig01}
\end{figure}

Second, perpendicular to the tonotopic axis, cells are arranged in
alternating bands according to binaural properties: bands of cells are
alternatively excited or inhibited by stimulation of the ipsilateral ear
(the contralateral ear usually produces an excitatory response${}^2$). The
tonotopic and binaural dominance organization is analogous to the
retinotopic and ocular dominance columns of visual cortex. Other parameters
have also been used to describe characteristics that change systematically
along isofrequency lines. Using combinations of two pure tones, one can
measure the Response Area (RA), also known as frequency-threshold curve,
i.e. the response threshold of a cell as a function of the tone frequency
presented. It has been shown that most RAs are topographically organized
along the isofrequency lines according to the symmetry of their excitatory
and inhibitory sidebands${}^3$. Other parameters have been also been shown
to change systematically in cat, such as threshold${}^4$, bandwidth${}^5$
and frequency modulation direction selectivity${}^{3,6}$.

These properties of AI cells are derived using pure tones (or clicks)
akin to using dots of light (or flashes) to study cells in the visual
pathway. Below we explain how to use the auditory version of drifting
gratings${}^7$ to characterize response properties of cells to dynamic
broadband sounds. This is necessary to gain insight to how timbre is
encoded. Another advantage of the method presented here is that it allows
us to determine the temporal and spectral properties of a cell at the
same time. In particular, one can study whether and to what extent the
response field varies as a function of time, thereby characterizing the
cell with a full spectro-temporal response field.

\section{Background}

\subsection{Response Field}

Traditionally, cells along the auditory pathway have been characterized
by their RA, or tuning curve. Determined using pure tones and by modifying
the frequency of the stimulus while adjusting its intensity, the RA is the
frequency-intensity combinations that elicit a threshold response, whether
the sustained activity level or the strength of the onset response. In this
paper, we use the Response Field (RF), a function measured using broadband
sounds. As illustrated in Figure~\ref{fig02}, it roughly reflects the
range of frequencies that influence the discharge properties of the neuron
under study. It is given in the form of a function, with positive values
describing excitation (proportional to the RF's amplitude) and negative
values describing inhibition. In general, the RF is a spectro-temporal
function, as opposed to the RA which typically describes only static
properties (but see Nelken${}^8$ et al and Sutter${}^9$ et al). The
definition of RF will be made more precise later.

\begin{figure}
\centerline{
\epsfbox{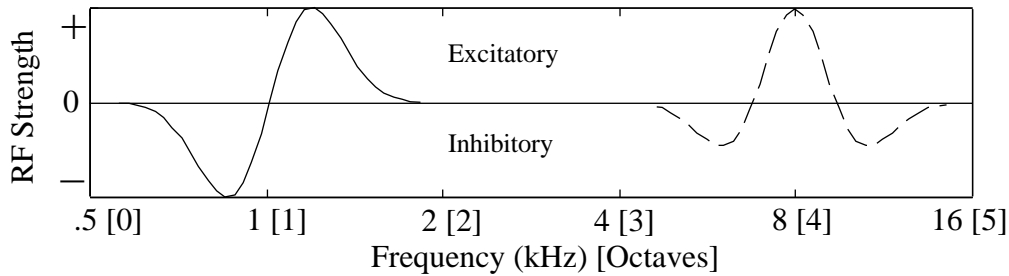}}
\caption[Examples of idealized RFs]
{\small\spacing{1}\sf Two idealized RFs at a given time. One RF (unbroken
line) is centered on low frequencies and is asymmetric, and the other
(broken line) is centered on high frequencies and is symmetric.}
\label{fig02}
\end{figure}

\subsection{Natural Sounds}

Natural sounds, such as environmental sounds, music and speech, are
classified along several perceptual axes. We typically describe a sound
by its loudness, its pitch and its timbre. Pitch is what changes when we
pronounce the same vowel with different tonal heights, e.g. the pitch of a
female voice is typically higher than the pitch of a male voice. Timbre is
what changes when, keeping the same tonal height, we pronounce different
vowels (e.g. /ah/, /eh/, /ih/). Figure~\ref{fig03} illustrates the
spectral profile or envelope of a sound. The envelope of a sound can
be viewed as a low-order polynomial fit of the (time-windowed) spectrum
of the sound. A common method for the extraction of the envelope is the
Linear Predictive Method (LPC)${}^{10}$; we will not go into the details of LPC
here, instead referring the reader to the intuitive notion of envelope
illustrated in Figure~\ref{fig03}.

\begin{figure}
\centerline{
\epsfbox{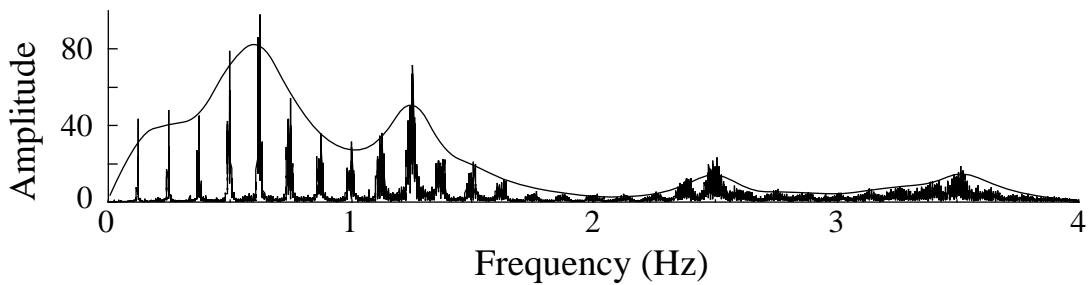}}
\caption[Spectral envelope of a vowel]
{\small\spacing{1}\sf The spectrum of /aa/ spoken by one of the authors,
with the spectral envelope superimposed on it.}
\label{fig03}
\end{figure}

The percept of timbre has been typically ascribed to the extraction of
the envelope of the spectrum, but it also includes the temporal variations
in the spectral envelope (for instance, the sound of a piano note played
backwards sounds more like that of a wind organ, even though the amplitude
of the Fourier transform of a sound and its time-reversed version are
identical). Therefore, the study of how timbre is encoded must include
temporal as well as spectral properties of the system. For speech, the
temporal variations in timbre involve time-scales of about 10 Hz, so that
this dimension of time is different from the temporal frequencies that
make up sounds. It is the extraction of the dynamic spectral envelope by
the auditory cortex that we are concerned with. Because we are interested
in timbre, we use pitchless, dynamic, broadband sounds as stimuli.

\subsection{Auditory Pathway (Monaural)}

The auditory pathway up to primary auditory cortex, ignoring structures
usually considered dedicated to binaural aspects of sounds (such as
localization) can be minimally described as follows. The vibrations of the
tympanic membrane are mechanically transformed into a traveling wave in
the cochlea, with a profile that depends on the frequency content of the
acoustic spectrum. The vibrations of the basilar membrane are transformed
by inner hair cells into patterns of neural activity in the auditory
nerve. For practical purposes, we can think of the basilar membrane as
a collection of 1/3 octave filters, performing a time-windowed Fourier
transform, with a time characteristic of about 30 ms. The auditory nerve
projects to the Cochlear Nucleus, which contains a variety of cells with
different properties. These cells project to the Lateral Lemniscus, then
to the Inferior Colliculus, then to the Medial Geniculate Body in the
Thalamus, and finally to the Auditory Cortex. As with all other sensory
modalities, there are strong back projections for most forward projections.

Neurons at different stages of the auditory pathway respond to different
time-scales. Neurons in the mammalian auditory nerve phase-lock to a
pure tone up to frequencies of about 4 kHz: that is, they tend to fire
at a specific phase of the tonal input, even if they fire in a sustained
fashion at the maximum rate of about 200 spikes/second.${}^{11}$ In the
cochlear nucleus certain cells (so-called lockers) can phase-lock to tones
for frequencies up to about 2 kHz.${}^{12,13}$ By the Inferior Colliculus,
most cells phase-lock to variations in the stimulus up to about 200 Hz
with some cells going up to 800 Hz.${}^{14,15}$ Finally, at the level of
cortex, we have found that phase-locking to variations in the stimulus
is usually on the order of 10 Hz with a maximum of about 70 Hz.${}^{16}$
Characterizing single units and their temporal features may ignore other
potential coding strategies based on population activity. In the cat's
cochlea, 3000 inner hair cells innervate 50,000 auditory fibers,${}^{17}$
and by the auditory cortex, activity has been distributed over several
millions of neurons.

Another important aspect of the organization of the auditory pathway is
that cells tend to be organized in a tonotopic manner at each step: the
frequency decomposition performed by the basilar membrane is along an
axis which is logarithmic. Up through AI, cells that are equally spaced
along a certain axis (which depends on the structure) respond best to
sounds that are linearly spaced on a logarithmic frequency axis.

\section{Principles}

\subsection{Guiding Principles}

The guiding principle behind our research program is that cells behave
like a linear system with respect to the spectral envelope. The proof of
linearity is that when cells are presented with a sound made of up the sum
of several spectral envelopes, the response, as measured assuming a rate
code, is the sum of the responses to the individual envelopes. A response
linear in frequency and time is characterized by a two-dimensional impulse
response (or time-dependent response field) or equivalently, its Fourier
transform, a two-dimensional transfer function. The extraction of this
two-dimensional response field, a function of frequency and time, is the
object of this paper.

It is helpful to remember that because the cochlea performs in some
sense a time-windowed Fourier transform of the incoming waveform along
its length, it is constructive to treat the frequency axis as a spatial
axis, not the Fourier transform of the time axis. Since the frequencies
are mapped logarithmically along the cochlear axis, the natural unit along
the spectral axis is $x = \log(f)$. Much research on which the present
work is based has dealt with the spectral, time-independent aspect of
the response fields and linearity.${}^{18}$

\subsection{Response Field and Linearity}

Initially ignoring the dimension of time (or taking a delta function
for the temporal impulse), the response of a cell with a response field
$RF(x)$, to a sound with a spectral envelope $S(x)$, is given by
$y=\smallint S(x)\cdot RF(x)dx$.\footnote{This is the standard convention used
in hearing and vision in defining the Response Field; it is related to the
Spectral Impulse Response function, which is $RF(-x)$.} Incorporating time
(or allowing for more realistic temporal Impulse Response functions),
we first limit our study to the case in which the temporal and spectral
properties that characterize the cells' responses are independent one
from the other (separable). The response of a cell is then characterized
by two functions, $RF(x)$, which describes the spectral properties, and
$IR(t)$, which describes the temporal properties of the cell. Then, the
response of a cell is described by
$y(t)=\left( {\smallint S(x,t)\cdot RF(x)\;dx} \right)*IR(t)$
where $*$ is the convolution operator. We will see that we can
characterize certain cells in this way.

In the general situation, cells must be characterized by a full
spectro-temporal description, i.e. a Spectro-Temporal Response Field,
$STRF(x,t)$. In this case the response is given by $y(t)=\smallint
S(x,t)*_t STRF(x,t)\;dx$, where the $*_t$ means convolution in the $t$
direction (with multiplication in the $x$ direction).

In the following, it is useful to consider the Fourier transform of
the two-dimensional impulse response function, $STRF(-x,t)$, called the
transfer function, $T\left( {\Omega,w} \right)$, where we define
$T\left( {\Omega ,w} \right)={\cal F}_{\Omega ,w}\left[ {STRF(-x,t)} \right]$.
The coordinate dual to $x$ is $\Omega$, and
the coordinate dual to $t$ is $w$\footnote{The coordinate dual to $t$ is w, not
f. This is because the spectro-temporal representation we are using is
inspired by the cochlea's time-windowed Fourier transform on the original
(acoustic) input signal. The time coordinate $t$ used at higher levels in
the auditory pathway is much coarser than the acoustic time, roughly
corresponding to a labelling of "which" cochlear time-window is being
referred to.}.

\subsection{Spectro-Temporal Response Field}

Our general problem can be formulated as follows: $S(x,t)$ is the
spectro-temporal envelope of the sound. Given the $STRF(x,t)$ of a neuron,
we can measure its response to any $S(x,t)$. We obtain this $STRF$ from
measurements of the neuron's response to a complete set of basis functions
$S_{\Omega w}\left( {x,t} \right)$. A simple set of basis functions is
$S_{\Omega w}(x,t)=\sin 2\pi (\Omega \cdot x+w\cdot t)$ where $S =
0$ corresponds to a flat envelope of fixed loudness (i.e. noise). Any
orthogonal basis will do, but the use of a sinusoidal basis allows us
to use the standard methods of Fourier analysis. Furthermore, because
of non-linearities discussed below, the sinusoidal basis is robust
against distortion. We use the sinusoidal basis functions, and call
them `ripples'. For this reason $\Omega$ is called ripple frequency
(in cycles/octave) and w is called ripple velocity (in cycles/second,
or Hertz).

\begin{figure}
\centerline{
\epsfbox{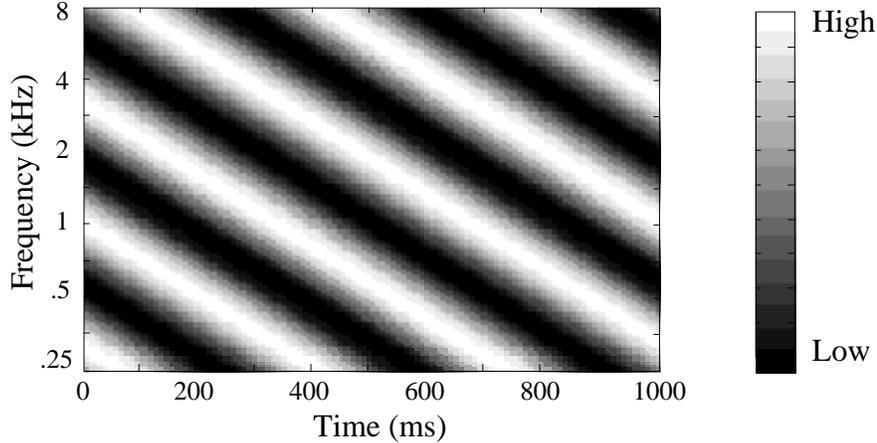}}
\caption[Spectro-temporal envelope of a ripple]
{\small\spacing{1}\sf
Spectrotemporal envelope of a ripple, moving downward in frequency with
$w$ = 3 Hz and $\Omega$ = 0.6 cycles/octave.}
\label{fig04}
\end{figure}

The most prominent non-linear distortions are half-wave rectification and
compression. The half- wave rectification is due to the impossibility of
negative spike rates (assuming the steady-state response to a flat spectrum
to be zero, as will be seen to be the case); the distortion of a sinusoid
due to firing rate half-wave rectification does not affect the phase of the
response, and its effect on the amplitude of the first Fourier component
is a constant factor (independent of $\Omega$ and $w$). The distortion
due to compression does not affect the phase of the response.

\subsection{Transfer Functions}

By measuring the response $y_{\Omega w}\left( t \right)$ of a cell to
a ripple of specific ripple frequency $\Omega$ and ripple velocity $w$,
we can obtain the transfer function $T\left( {\Omega ,w} \right)$ at one
point in $\Omega -w$ space.

\begin{eqnarray}
&y_{\Omega w}\left( t \right)&=\int\!\!\!\int {dx'dt'\,STRF\left(
{x',t'} \right)\ \sin 2\pi \left( {\Omega x'+w\left( {t-t'} \right)}
\right)}\nonumber\\
 &&=\Im \,\int\!\!\!\int {dx'dt'\;STRF(x',t'){\rm e} ^{2j\pi \left( {\Omega
x'+w\left( {t-t'} \right)} \right)}}\nonumber\\
 &&=\Im \left[ {{\rm e} ^{2j\pi wt}\int\!\!\!\int {dx'dt'\ }STRF(x',t'){\rm
e}^{2j\pi \left( {\Omega x'-wt'} \right)}} \right]\nonumber\\
 &&=\Im \left[ {{\rm e} ^{2j\pi wt}\,\,F_{\Omega ,w}\left[ {STRF\left( {-x',t'}
\right)} \right]} \right]\nonumber\\
 &&=\Im \left[ {{\rm e} ^{2j\pi wt}\,\,T\left( {\Omega ,w} \right)} \right]\nonumber\\
 &&=\Im \left[ {{\rm e} ^{2j\pi wt}\,\,\left| {T\left( {\Omega ,w} \right)}
\right|{\rm e} ^{j\Phi \left( {\Omega ,w} \right)}} \right]\nonumber\\
 &&=\left| {T\left( {\Omega ,w} \right)} \right|\ \sin \left[ {2\pi
wt+\Phi \left( {\Omega ,w} \right)} \right]
\end{eqnarray}

In this way, we derive the amplitude $\left| {T\left( {\Omega ,w} \right)}
\right|$ and phase $\Phi \left( {\Omega ,w} \right)$ of the complex
transfer function $T(\Omega,w)$ by measuring the amplitude and phase of the
(real) response of the cell. By the definition of the transfer function,
it follows that the inverse Fourier transform of $T(\Omega,w)$ is the
STRF of the cell:
$STRF(x,t) = {\cal F}^{-1}{}_{-x,t}\left[T_{\Omega w}\right]$.

Because $STRF(x,t)$ is real, but $T(\Omega,w)$ is complex, there
is a complex conjugate symmetry,
\begin{equation}
T\left( {\Omega ,w} \right)=T^*\left( {-\Omega ,-w} \right)
\end{equation}
which holds for the Fourier transform of any real function of $x$ and $t$.

\begin{figure}
\centerline{
\epsfbox{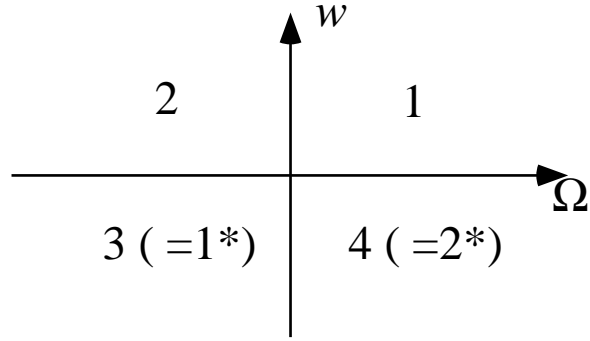}}
\caption[The $w$-$\Omega$ plane]
{\small\spacing{1}\sf
The $\Omega$ - $w$ plane. The value of the transfer function at a point in
quadrant 1 is the complex conjugate of the value at the corresponding
reflected point in quadrant 3 (and similarly for the quadrant pair 2 \&
4). The ripple in Figure~\ref{fig04} corresponds to a pair of points in
quadrants 1 and 3.}
\label{fig05}
\end{figure}

\subsection{Full Separability}

Many cells possess transfer functions that are fully separable, i.e. the
ripple transfer function factorizes into a function of $\Omega$ and a
function of $w$ over all quadrants: $T(\Omega,w) = F(\Omega)\cdot G(w)$.
This implies that $STRF(x,t)$ is spectrum-time separable: $STRF(x,t) =
RF(x) \cdot IR(t)$. In this case, we only need to measure the transfer
function for all $\Omega$ at an arbitrary $w$, and for all $w$ at an
arbitrary $\Omega$. Then $F(\Omega)$ and $G(w)$ are each complex-conjugate
symmetric (because $RF(x)$ and $IR(t)$ are real), and we need only consider
the positive values of each. This dramatically decreases the number of
measurements needed to characterize the STRF.

\subsection{Quadrant Separability}

For cells that are not fully separable, we have found that they are still
quadrant separable,${}^{16}$ i.e. the transfer function $T(\Omega,w)$
can be written as the product of two independent functions:
\begin{equation}
T\left( {\Omega ,w} \right)=\left\{ \matrix{F_1\left( \Omega
\right)\,G_1\left( w \right)&\Omega >0,w>0\hfill\cr
 F_2\left( \Omega \right)\,G_2\left( w \right)&\Omega <0,w>0\hfill\cr}
\right.
\end{equation}
where the subscript 1 indicates the $\Omega >0,w>0$ quadrant, and the
subscript 2 the $\Omega <0,w>0$ quadrant. Note that by reality of the
STRF, the transfer function in quadrants 3 ($\Omega <0,w<0$) and 4 is
complex conjugate to quadrants 1 and 2 respectively. In this case, the
STRF is {\em not} separable in spectrum and time, but is the linear
superposition of two functions, one with support only in quadrant 1
(and 3), and one with support only in quadrant 2 (and 4).

\subsection{Confirming Separability}

Separability is measured by comparing the measured transfer function
taken along parallel lines of constant $\Omega$ or constant $w$. If the
sections of the transfer function differ only by a constant amplitude
and phase factor, then that section is independent of the perpendicular
variable and therefore the transfer function is separable. If in addition
the section of the transfer function is complex-conjugate symmetric about
zero, then the transfer function is fully separable. Otherwise the transfer
function is merely quadrant-separable.

\subsection{Confirming Linearity}

The method we use to characterize cortical cells depends on their being
linear, so linearity must be assessed. To this end, we measure (as
described above) the transfer function of a cell with single ripples,
and then measure the extent to which we can predict the response of
the cell to a linear combination of ripples. Confirmation of linearity
comes from measuring the response of the cell to linear combinations
of ripples, thereby verifying the degree of linearity of the response.

Predicting the response of the cell to linear combinations of ripples for
which the transfer function was not measured directly, but only inferred
via separability, verifies both linearity and separability simultaneously.

\subsection{Characterizing the Response}

The functions $F(\Omega)$ and $G(w)$ are unconstrained
theoretically. Physiologically, however, there are constraints on the type
of functions they may be. For instance, because $F(\Omega)$ is the Fourier
transform of $RF(x)$ which is localized around a center frequency ($f_m$
in frequency space, $x_m$ in logarithmic frequency space), the phases of
$F(\Omega)$ must constructively interfere at $x_m$, and the amplitude of
$F(\Omega)$ must be band limited. See, e.g. Figure~\ref{fig02} for examples
of RFs, each of which is band limited and centered at a different $x_m$.

\subsubsection{Amplitude of the response}

The amplitude of the ripple frequency transfer function $F(\Omega)$ reaches
a maximum at $\Omega _m\approx \left( {2BW} \right)^{-1}$, where $BW$
is the excitatory bandwidth of the RF in octaves, and then decreases: at
higher ripple frequencies the modulations of the ripple's spectral envelope
cancel when integrated against the (more slowly varying) RF; at ripple
frequencies lower than $\Omega_m$, the energy in the ripple's spectrum is
fairly constant over the width of the RF, including any negative sidebands,
and therefore integrates to a smaller magnitude. Similarly, the amplitude
of the ripple velocity transfer function G(w) has a maximum at $w_m\approx
\left( {2BW_t} \right)^{-1}$, where $BW_t$ is the temporal excitatory
width of the IR. Because under anesthesia the steady state response to
any sound with a constant envelope has a rate of zero in cortex, we get
$G(0)=\int {dt\,IR\left( t \right)}=0$.

\subsubsection{Phase of the response}

Because neurons in the auditory pathway are tonotopically arranged,
each cell has a frequency around which the RF is centered which is
independent of the ripple frequency $\Omega$. Since the derivative of the
phase of $F(\Omega)$ gives the mean frequency of the response for that
ripple frequency, the phase of the transfer function is linear (plus
a constant)\footnote{This is completely analogous to the derivative of
the phase of the Fourier transform of a signal, $d\phi /dw$, giving the
characteristic delay (for that frequency) or the derivative of the angular
frequency of a dispersion relation, dw/dk, giving the group velocity
(for that wave number). See, e.g. Papoulis${}^{19}$ and Cohen${}^{20}$.}.
Similarly, because IR is causal, there is a group delay, and because
of the biological nature of the neural process, the delay is roughly
independent of ripple velocity, which gives a constant derivative of the
phase of $G(w)$.

Therefore the phase of the transfer function $\Phi ^q\left( {\Omega ,w}
\right)$ (see Equation~(1)), $q=\left\{ {1,2} \right\}$ (for each quadrant),
can be written as $\Phi ^q\left( {\Omega ,w} \right)=2\pi \Omega x_m^q+2\pi
w\tau _d^q+\chi ^q$, where $x_m^q=\log f_m^q$ is the mean frequency around
which the RF is centered, and $\tau _d^q$ is the delay of the IR, defined
as the mean of the envelope of the IR.\footnote{The envelope $E(t)$ of
a function with localized support can be defined as the modulus of the
function plus $j$ times its Hilbert transform. The mean of the envelope is
then computed as $\left\langle t \right\rangle =\int {dt\,t\,E(t)^2}$. See,
e.g. Cohen${}^{20}$.}. $\chi ^q$ is a constant phase angle. Tonotopy
guarantees that $x_m^1\approx x_m^2$, but depending on the precise inputs
of the neuron, they may not agree completely, so that we can have different
$x_m$ for upward and downward moving sounds. Similarly, $\tau _d^1\approx
\tau _d^2$, but equality is not required. The reality of the response
enforces complex-conjugate symmetry of the transfer functions, allowing
for these six independent parameters to describe the phase everywhere in
the $\Omega -- w$ plane. A convenient convention is to define constant
phase angles $\theta$ and $\phi$ such that $\chi ^1=\theta +\phi ,\;\chi
^2=\theta -\phi $. With the complex-conjugate symmetry, and if the STRF
is separable, $\phi$ is the symmetry parameter of the RF and $\theta$
is the symmetry parameter of the IR (in Figure~\ref{fig02}, $\phi =
90^o$ for the left cell and $\phi = 0^o$ for the right cell). Even in
the non-separable case, we will still call $\phi$ the RF symmetry and
$\theta$ the IR polarity. If one restricts measurements to one quadrant
plus the $w$-axis (recall from above that the transfer function vanishes
on the $\Omega$-axis), one can measure $\chi$ in that quadrant and, on
the axis, the average of $\chi^1$ and $\chi^2$, i.e. $\theta$. There is
an ambiguity in fixing $\theta$ and $\phi$ that allows us to restrict
$\theta$ to lie between $0^o$ and $180^o$, while $\phi$ ranges the full
$-180^o$ to $+180^o$.

\begin{figure}
\centerline{
\epsfbox{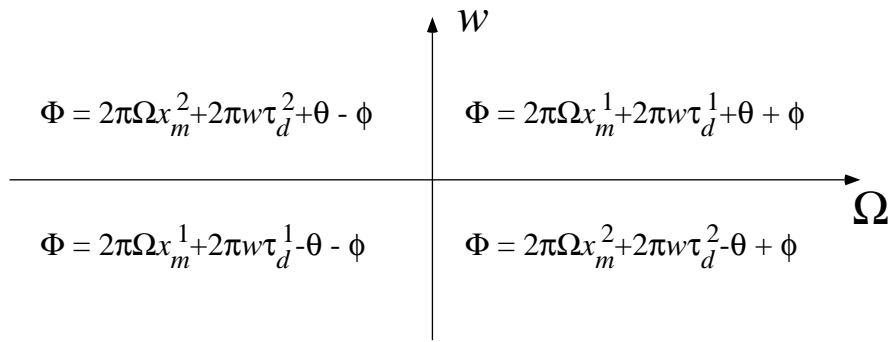}}
\caption[Characterizing the phase of the transfer function]
{\small\spacing{1}\sf
The phase of the transfer function can be described by 6 parameters
over most of the relevant regions of the $\Omega$ - $w$ plane.}
\label{fig06}
\end{figure}

The phase curve does not truly have a discontinuity across the axis. For
very small ripple frequencies, the response becomes more independent of
the best frequency of the cell, allowing the slope to change continuously
from its constant value to $\theta$. At large ripple frequency the slope may
also diverge from its constant value, but at these ripple frequencies
the amplitude is small and so the particular values of the phase do not
contribute. Similarly, the phase of $G(w)$ is constant over its intermediate
range but changes continuously to $\phi$ on the $\Omega$-axis. Since the
amplitude is zero on that axis, this is not so important.

\begin{figure}
\centerline{
\epsfbox{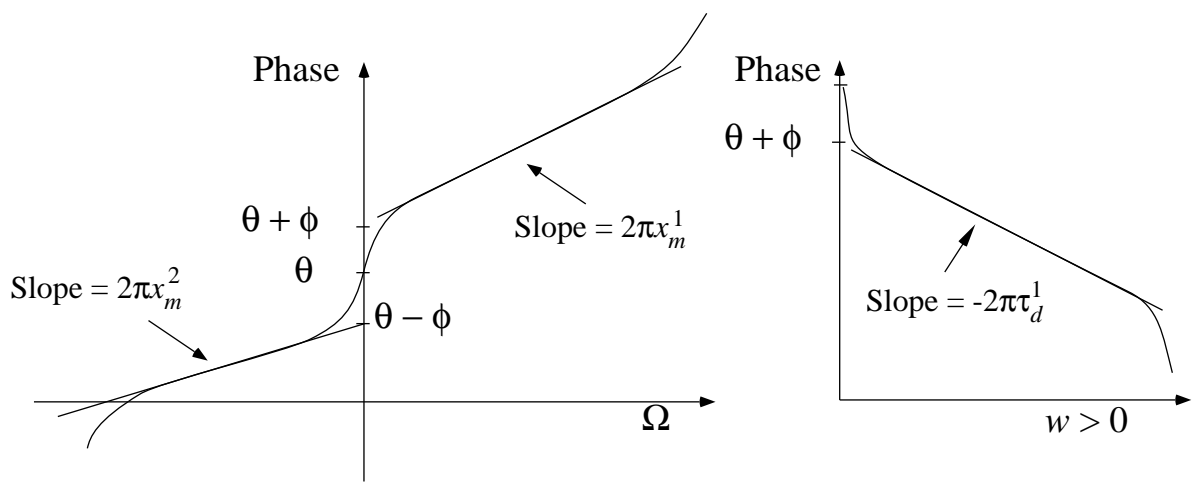}}
\caption[Phase curves]
{\small\spacing{1} \sf
Phase Curves. The slope is constant for most of the curves, after
(left)
$2\pi w\tau _d^q$
has been removed from the corresponding quadrants, corresponding
to a center frequency that is independent of the ripple frequency,
and (right) after
$2\pi \Omega x_m^1$
has been removed, corresponding to a delay that is
independent of ripple velocity. At very small ripple frequencies (long
ripple periodicity), center frequency is less meaningful, and similarly for
small ripple velocity and delay, respectively. At large ripple velocity
the slope asymptotes to the signal-front delay, but when this occurs the
small amplitude of the transfer function makes it difficult to measure
the phase (see Dong and Atick${}^{21}$ and Papoulis${}^{19}$).}
\label{fig07}
\end{figure}

\section{Analytical Methods}

\subsection{The Ripple Stimulus}

The auditory stimulus we use has a sinusoidal profile at any instant
in time. Since it would be hard to generate noise and then shape it
with filters, we generate ripples over a range of 5 octaves by taking
101 tones with logarithmically spaced (temporal) frequencies and random
(temporal) phases. The amplitude $S(x,t)$ of each tone of frequency $f$,
with $x=\log_2(f\/ f_0)$, $f_0$ the lower edge of the spectrum, is then
adjusted as
\begin{equation}
S(x,t)=L\left( {1+\Delta A\cdot \;\sin \left(
{2\pi \left( {\Omega \cdot x+w\cdot t} \right)+\Phi } \right)} \right),
\end{equation}
for a linear modulation. $L$ is the overall base of the stimulus and is
adjusted to a level typically 10-15 dB above the lower threshold of
the cell as determined with pure tones at the tonal best frequency. The
overall level of a single-ripple stimulus is calculated from the level
of its single frequency components: thus, a flat ripple of level $L_1$ is
composed of 101 components, each at
$L_1\,-10\;\log \left( {101} \right)\approx L_1\,-20\;dB$.

\begin{figure}
\centerline{
\epsfbox{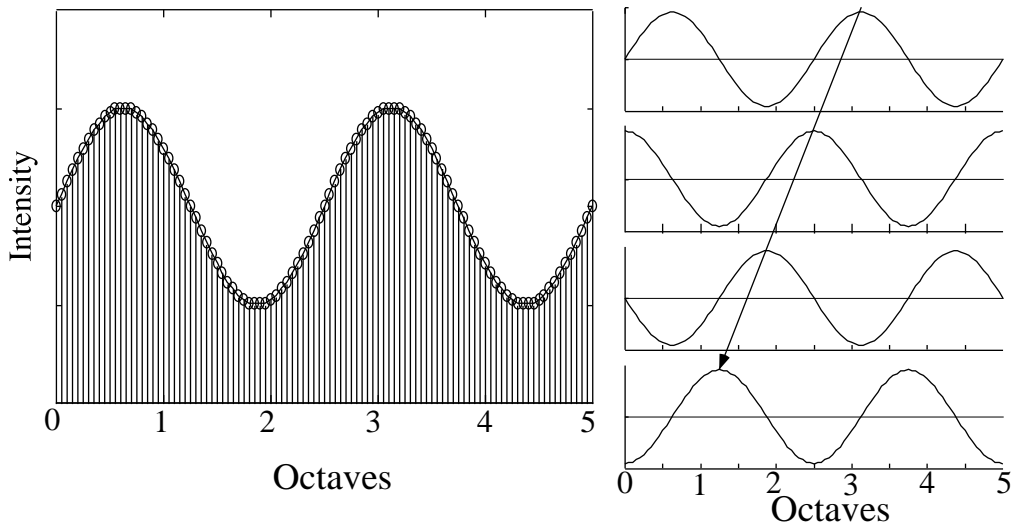}}
\caption[Generating the ripple stimulus]
{\small\spacing{1} \sf
Left: A time slice of the stimulus: 101 tones equally spaced along the
logarithmic axis. This ripple has a ripple frequency $\Omega$ of 0.4 cyc/oct
with zero phase, and a linear modulation of 50\%, against an arbitrary
intensity axis (see Equation~(4)).
Right: the spectral profile changes as a function
of time, giving a moving ripple, here with positive frequency (since the
phase increases as a function of time). }
\label{fig08}
\end{figure}

Five parameters are sufficient to characterize the ripple stimulus:
\begin{itemize}
\item The ripple frequency $\Omega$ in cycles/octave,
\item The ripple velocity $w$ in Hz, so that a positive value of $w$ and
$\Omega$ corresponds to a ripple whose envelope travels towards the
low frequencies
\item The level or base loudness of the ripple,
\item The amplitude of the modulation $\Delta A$ of the ripple around the base,
\item The ripple's initial phase.
\end{itemize}

Since the tones that make up a ripple are logarithmically spaced, its
pitch is indeterminate.

\subsection{Data Analysis}

In this section, we show the data analysis we apply with the help of
a simulation, but to keep the graphs one-dimensional we assume that in
Figure~\ref{fig09} and Figure~\ref{fig10},
$IR\left( t \right)=\delta \left( t \right)$.

\begin{figure}
\centerline{
\epsfbox{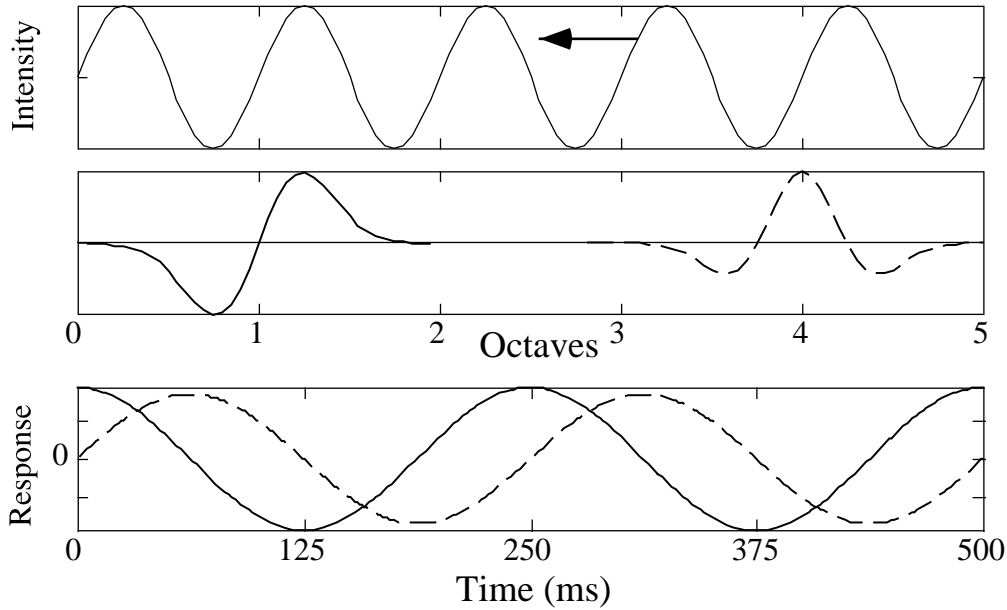}}
\caption[Schematic of the response]
{\small\spacing{1} \sf
The top panel represents the spectral envelope of the stimulus at a
given instant against an arbitrary intensity axis. For the two cells
represented in the middle panel (with
$IR\left( t \right)=\delta \left( t \right)$), one (unbroken) with the RF
centered on low frequencies ($x_m = 1$, asymmetric with $\phi=90^o$),
and the other
(broken) with the RF centered on high frequencies ($x_m = 4$, symmetric with
$\phi = 0^o$), the expected responses to a 4 Hz ripple is shown
in the bottom panel
(unbroken and broken, respectively),against some measure of the response,
for instance spikes/sec or the intracellular potential. In our case, the
actual response is half-wave rectified, and measured in the form of a
spike count, so that the bottom panel should really be seen as a spiking
probability that can be measured by measuring the response of the cell to
many presentations of the same stimulus.}
\label{fig09}
\end{figure}

We use two paradigms to obtain the transfer function of a cell. First, we
choose a ripple frequency and present the cell with ripples of varying
ripple velocities (typically, -24 Hz to 24 Hz in cortex). Then, for a
fixed ripple velocity, we present the cell with ripples of varying ripple
frequencies (typically, from -1.6 to 1.6 cyc/oct).

As indicated
for a 4 Hz ripple in Figure~\ref{fig09}, the response of a cell as a
function of time is modulated at the same (temporal) frequency as that
of the stimulus. Therefore, we just have to extract the phase and the
amplitude of the response. The resulting transfer function for the same
two cells is shown in Figure~\ref{fig10}. We have presented ripples to
the idealized cells shown in panel B. The amplitude of the response as
a function of ripple frequency is shown in panel C, whereas the phase of
the response is shown in the bottom panel. Note that the phase intercept
$\phi$ is $0^o$ for the symmetric cells and $90^o$ for the antisymmetric cell.

\begin{figure}
\centerline{
\epsfbox{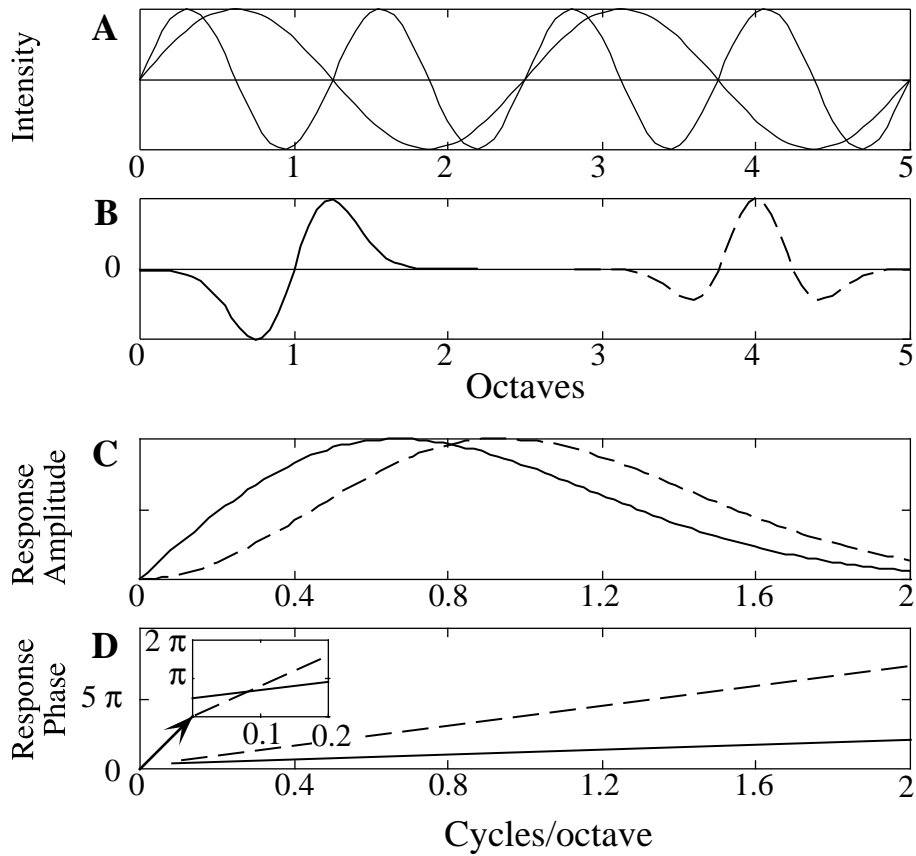}}
\caption[Computing the transfer function]
{\small\spacing{1} \sf
The sounds with the spectrum shown in A (ripples with ripples frequencies
of 0 (flat spectrum), 0.4 and 0.8 cycles/octave) are presented at various
phases to the two cells in B, as in Figure~\ref{fig09}.
The amplitude (for instance in
spikes/sec) (C) and phase (D) of the best fit to the response are shown.}
\label{fig10}
\end{figure}

In the corresponding $\Omega--w$ space, the ripple of Figure~\ref{fig08}
corresponds to a pair of points. Therefore, to measure the complete ripple
response transfer function of a cell we need to measure its response to
all possible ripples, as shown in Figure~\ref{fig11}. Note that since
cells in cortex respond only to transient stimuli, it is not necessary
to present the stimuli along the $w=0$ axis.

\begin{figure}
\centerline{
\epsfbox{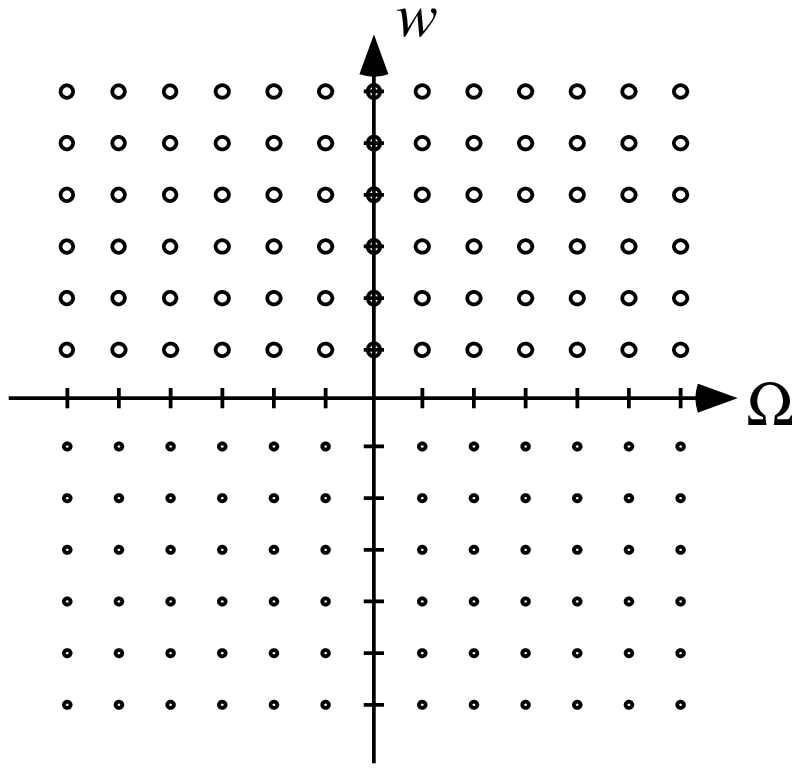}}
\caption[Computing the transfer function: worst-case scenario]
{\small\spacing{1} \sf
To measure the complete ripple transfer function, we have to measure
the response of the cell to all the ripples represented by large circles
above. The smallest circles correspond to redundant ripples, as inspection
of Eq.~(2) and Figure~\ref{fig05} shows.}
\label{fig11}
\end{figure}

\subsection{Separability}

We have shown previously${}^{16,22}$ that within each quadrant, actual
ripple transfer functions are separable \footnote{Strictly speaking, we
have shown it only for the first quadrant, i.e. for down-moving ripples.}:
for two fixed values of $\Omega$, the transfer function as a function of
$w$ only changes between the two by an overall scale factor and an overall
phase. The same is true when $\Omega$ and $w$ are reversed. Hence, one
is required only to study two lines in $\Omega -w$ space. Therefore we
only need to sample a line in each direction within each quadrant, as
shown in Figure~\ref{fig12}.

\begin{figure}
\centerline{
\epsfbox{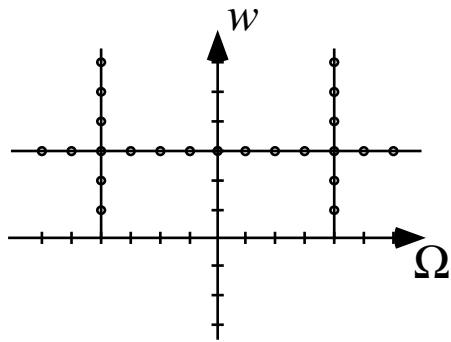}}
\caption[Computing the transfer function: actual scenario]
{\small\spacing{1} \sf
Since we found experimentally that cells have separable transfer
functions within each quadrant, it is enough to measure the transfer
function along two orthogonal lines in each quadrant.}
\label{fig12}
\end{figure}

Without separability, whether full or quadrant, it would be extremely
difficult to characterize a cell by its transfer function. Experimentally,
given the time required to measure one point of the transfer
function, measuring the transfer function at the points indicated in
Figure~\ref{fig12} is feasible, whereas measuring the transfer function
at all the points indicated in Figure~\ref{fig11} is not.

\subsection{Linearity}

Linearity is confirmed by comparing the response to combinations
of ripples with the response predicted by summing the responses to
the individual ripples, i.e. the values of the transfer function. A
combination of ripples is computed such that its base loudness is the
same as the individual ripples', and the amplitude of the modulation is
scaled as in Equation~(4). As an example, to present the combination of
two ripples (whose properties are described by subscripts 1 and 2), we
compute $B=B_1\;\sin \left( {2\pi \left( {\Omega _1\cdot x+w_1\cdot t}
\right)+\Phi _1} \right)+B_2\;\sin \left( {2\pi \left( {\Omega _2\cdot
x+w_2\cdot t} \right)+\Phi _2} \right)$. For a modulation of $\Delta A$,
the envelope is (in the manner of Equation~(4)) $L\,\cdot \left( {1+\Delta
\kern 1pt A\cdot {B \mathord{\left/ {\vphantom {B {\max \left( B \right)}}}
\right. \kern-\nulldelimiterspace} {\max \left( B \right)}}} \right)$,
where $L$ is the base intensity level. The sound is generated from the
envelope using 101 tones over 5 octaves with logarithmically spaced
(temporal) frequencies and random (temporal) phases.

\section{Experiment and Results}

\subsection{Experimental Details}

Data were collected from domestic ferrets (Mustela putorius). The ferrets
were anesthetized with sodium pentobarbital and anesthesia was maintained
throughout the experiment by continuous intravenous infusion of either
pentobarbital or ketamine and xylazine, with dextrose (in Ringer's
solution) to maintain metabolic stability. The ectosylvian gyrus, which
includes the primary auditory cortex, was exposed by craniotomy and
the dura reflected. The contralateral ear canal (meatus) was exposed
and partly resected, and a cone-shaped speculum containing a miniature
speaker was sutured to the meatal stump. For details on the surgery see
Shamma et al${}^{3}$.

All stimuli were computer synthesized, gated, and then fed through a common
equalizer into the earphone. Calibration of the sound delivery system (to
obtain a flat frequency response up to 20 kHz at the level of the eardrum)
was performed in situ using a 1/8-in. probe microphone.

Action potentials from single units were recorded using glass-insulated
tungsten micro- electrodes with 5-6 M$\Omega$ tip impedances. Neural
signals were fed through a window discriminator and the time of spike
occurrence relative to stimulus delivery was stored on a computer, which
also controlled stimulus delivery, and created raster displays of the
responses. In each animal, electrode penetrations were made orthogonal to
the cortical surface. In each penetration, cells were typically isolated
at depths of 350-600 $\mu$m corresponding to cortical layers III and
IV${}^{3}$.

\subsection{Obtaining the Transfer Functions}

As explained above, we measure the cells' transfer functions by presenting
first, at a fixed ripple frequency, ripples of various velocities. Then,
for a fixed ripple velocity, we present ripples of varying ripple
frequencies.

\subsubsection{Spectral cross-section of the transfer function}

A typical example of the analysis is shown in Figure~\ref{fig13}. Ripples
were presented at 8 Hz, for ripples frequencies from -1.6 cyc/oct to 1.6
cyc/oct in steps of 0.2 cyc/oct, with the ripple starting to move at $t
= 0 ms$, but being acoustically turned on starting at 50 ms with a linear
ramping over 8 ms. Each action potential is denoted by a dot on the raster
plot in A. One can see the onset response to the ripple at about 70 ms
(50 ms + delay due to the ramping up of the stimulus, + latency of the
response). Each ripple is presented 15 times. Once the onset activity
has died away, the cell goes into a sort of steady-state response. For
each ripple frequency, we compute a period histogram starting at 120 ms
(this excludes the onset response). Four of those histograms are shown in
panel B. To assess the strength and phase of the phase-locked response,
we divide the histogram into 16 equal bins. The amplitude and phase of
the response is then evaluated by performing a Fourier transform of the
data, and extracting the phase of
$T\left( {\Omega ,w=8\ Hz} \right)$ from the first component of the Fourier
transform, and the amplitude from
\begin{equation}
T\left( {\Omega ,w=8\ Hz} \right)=AC_1\left( \Omega \right)\cdot {{\left|
{AC_1\left( \Omega \right)} \right|} \over {\sqrt {\sum\nolimits_{i=1}^8
{\left| {AC_i\left( \Omega \right)} \right|^2}}}}
\end{equation}

If the modulation of the response were that of a purely linear
system, the higher coefficients $AC_i(\Omega)$ would be negligible. But because of the
half-wave rectification and other non-linearities, they usually are
significant. Therefore we weight $AC_1(\Omega)$ by the $RMS$ of the other coefficients
of the $AC_i(\Omega)$ to assess linearity.

The magnitude and phase of the transfer function is shown in panel C. In
D, we have inverse Fourier transformed separately the transfer function
in quadrant 1 and 2, or equivalently for down- and up-moving ripples,
after removing the constant (temporal) phase factor $2\pi w\tau _d+\theta
$, where $w = 8 {\rm Hz}$. In this case, the up- and down-moving RFs
match very well with each other and with the RF obtained with a two-tone
paradigm${}^{3}$.

Note that the period histograms shown in panel B correspond to periods
starting at 120 ms, so as to eliminate the effect of the onset response,
whereas the second graph in panel C shows phases sent back to 0 ms, at
which point in time the phase of the ripples presented were all 0 degrees.

\begin{figure}
\centerline{
\epsfbox{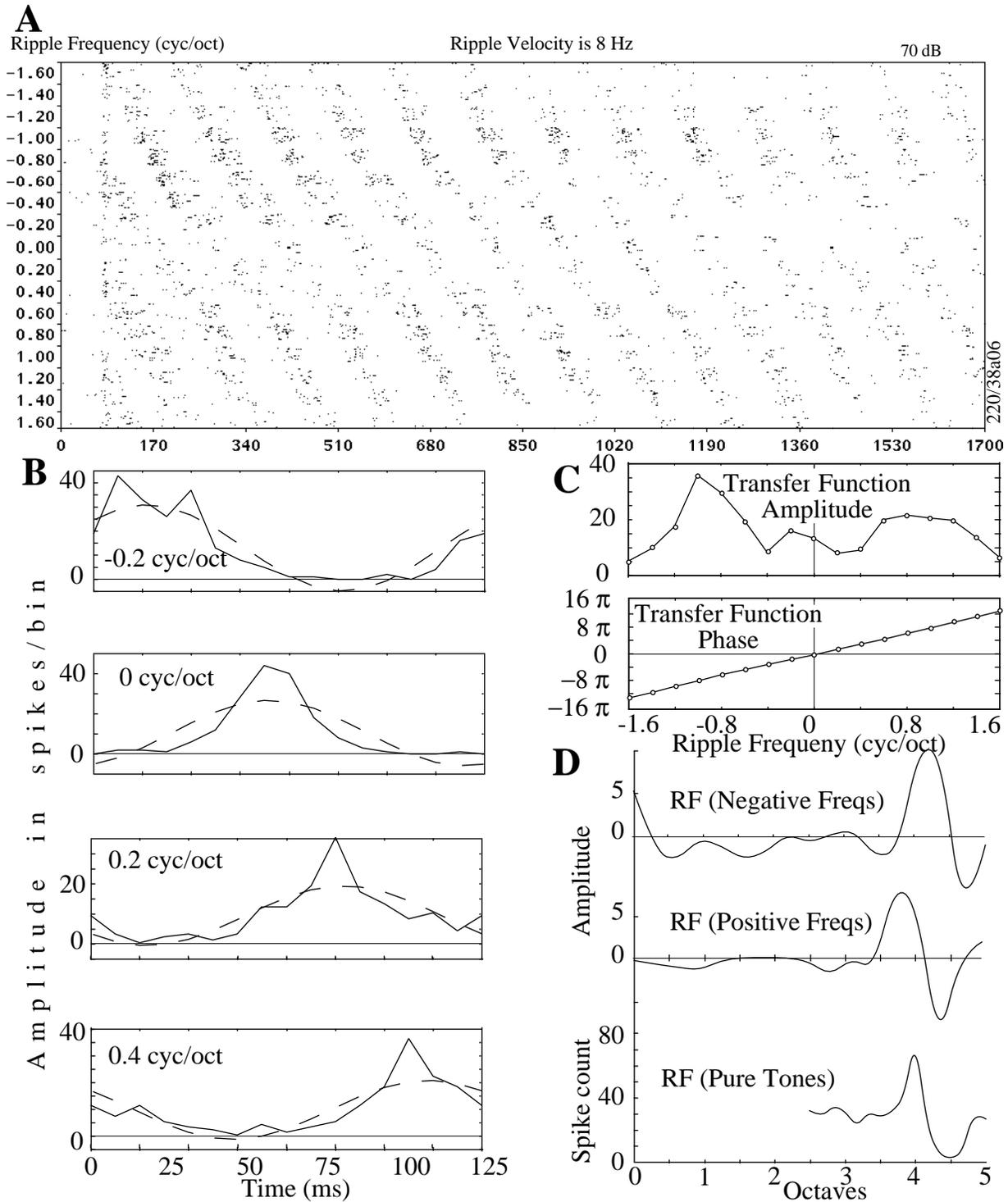}}
\caption[Measuring a spectral cross-section of the transfer function]
{\small\spacing{.9} \sf
Data analysis using ripples of fixed velocity and varying frequencies. A:
Raster plot of responses. Each point represents an action potential, and
each paradigm is presented 15 times. B: Period histogram for 4 ripple
frequencies. Note how the position of the peak of the best fit changes
linearly with ripple frequency. C: Magnitude and phase of the period
histogram fits. D: Separate inverse Fourier transforms for positive
and negative ripple frequencies of C, obtaining a slice of the RF. Also
given for comparison is the response area as determined by the two-tone
paradigm. ${}^{3}$}
\label{fig13}
\end{figure}

\subsubsection{Temporal Cross-Section of the Transfer Function}

An example of the extraction of the temporal cross-section of the
transfer function for the same cell as in Figure~\ref{fig13} is shown
in Figure~\ref{fig14}. Ripples are presented at 0.4 cyc/oct, for ripple
velocities from -24 Hz to 24 Hz in steps of 4 Hz, with the ripple starting
to move at $t = 0 ms$, being acoustically turned on starting at 50 ms
with a linear ramping over 8 ms. Each action potential is denoted by a
dot on the raster plot in A. One can see the onset response to the ripple
at about 70 ms (50 ms + delay due to the ramping up of the stimulus, +
latency of the response). Each ripple is presented 15 times. Once the
onset activity dies away, the cell goes into a steady-state response. For
each ripple frequency, we compute a period histogram starting at 120 ms
(so that the onset response is excluded). Four of those histograms are
shown in panel B. To assess the strength and phase of the phase-locked
response, we divide the period into 16 equal bins. The amplitude and phase
of the response is then evaluated by performing a Fourier transform of
the data, and extracting the phase of
$T\left( {\Omega =0.4\ cyc/oct,w} \right)$
from the first component of the
Fourier transform, and the amplitude from
\begin{equation}
T\left( {\Omega =0.4\ cyc/oct,w} \right)=AC_1\left( w \right)\cdot
{{\left| {AC_1\left( w \right)} \right|} \over {\sqrt {\sum\nolimits_{i=1}^8
{\left| {AC_i\left( w \right)} \right|^2}}}}
\end{equation}

If the modulation of the response were that of a purely linear system,
the higher coefficients $AC_i(w)$ would be negligible. But because of
the half-wave rectification and other non-linearities, they usually
are significant. Therefore we weight $AC_1(w)$ by the RMS of the other
coefficients of $AC_i(w)$ to assess linearity.

The magnitude and phase of the transfer function is shown in panel C. In
D, we have inverse Fourier transformed separately the transfer function
in quadrant 1 and 2, or equivalently for down- and up-moving ripples,
after removing the constant (spectral) phase factor
$2\pi \Omega x_m+\phi $, where
$\Omega =0.4\ cyc/oct$. In this
case, the up- and down-moving IRs match very well with each other.

\begin{figure}
\centerline{
\epsfbox{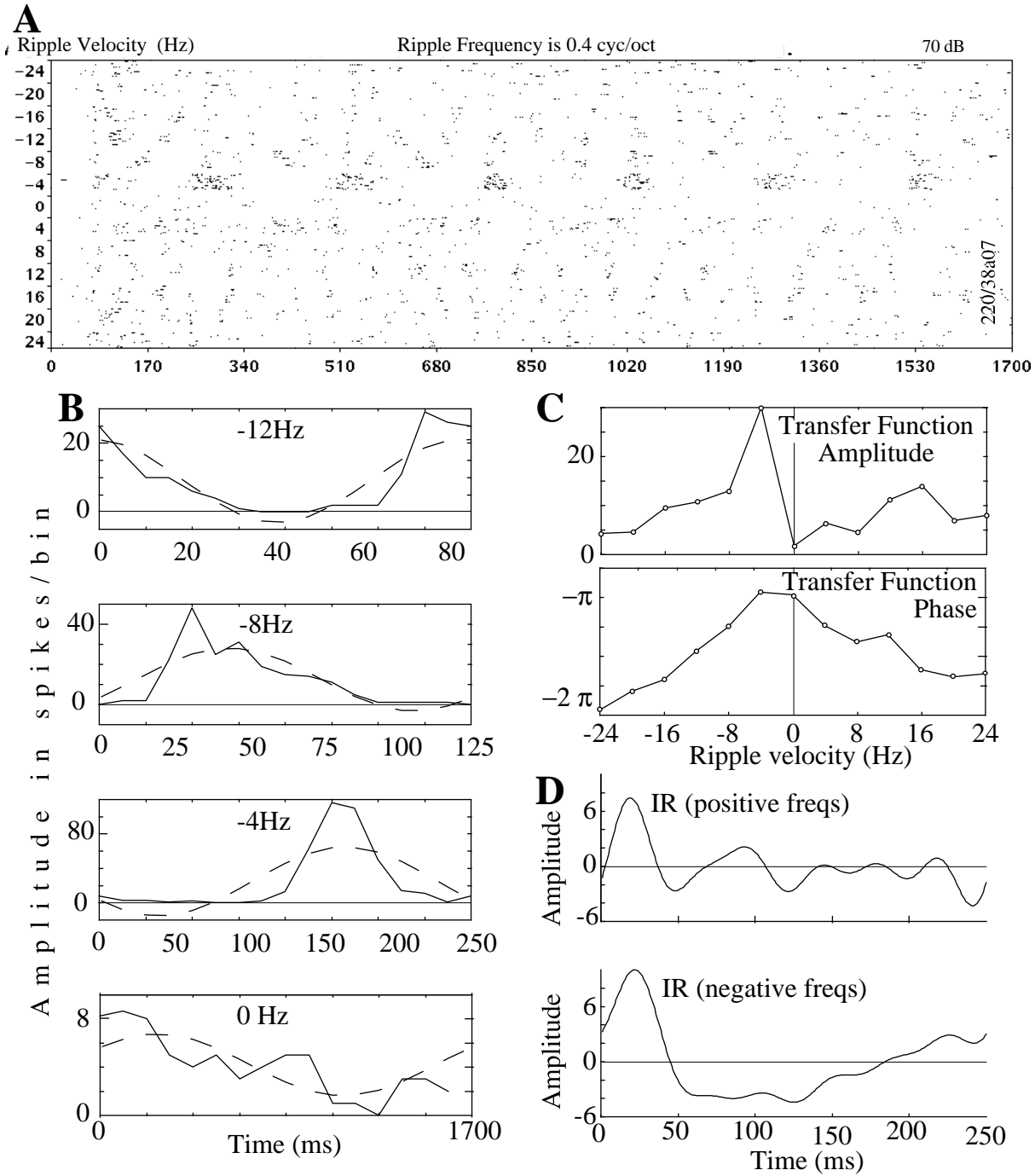}}
\caption[Measuring a temporal cross-section of the transfer function]
{\small\spacing{1} \sf
Data analysis using ripples of fixed frequency and varying velocities. A:
Raster plot of responses. Each point represents an action potential, and
each paradigm is presented 15 times. B: Period histogram for 4 ripple
velocities. Note how the peak of the best fit changes linearly with ripple
velocity (the 0 Hz case can be used to estimate noise). C: Magnitude and
phase of the period histogram fits. D: Separate inverse Fourier transforms
for positive and negative ripple velocities of C, obtaining a slice of
the IR.}
\label{fig14}
\end{figure}

\subsection{Quadrant Separability}

$RF(x)$ and $IR(t)$, as illustrated in panels D of Figure~\ref{fig13}
and Figure~\ref{fig14}, are linear combinations of the transfer function
evaluated along cross-sections of the $\Omega-w$ plane. Constancy of
$RF(x)$ computed for different $w$ is equivalent to proportionality
of $T\left( {\Omega ,w} \right)$ for different $w$ (and similarly for
$RF(x)$, $\Omega$, and $T\left( {\Omega ,w} \right)$). This was the
requirement given above to verify quadrant separability. This has all
been verified for many cells in the first quadrant${}^{16}$. While it
is theoretically possible for the remaining independent quadrant to
be nonseparable, it seems unlikely in ferrets, humans, and most mammals
(possible exceptions might include sonar-using animals, which could require
further specialization). We are currently verifying separability in the
second quadrant.

Shown in Figure~\ref{fig15} are examples of the positive-frequency RF
and positive-frequency IR for two cells, as computed at the different
sections indicated.

\begin{figure}
\centerline{
\epsfbox{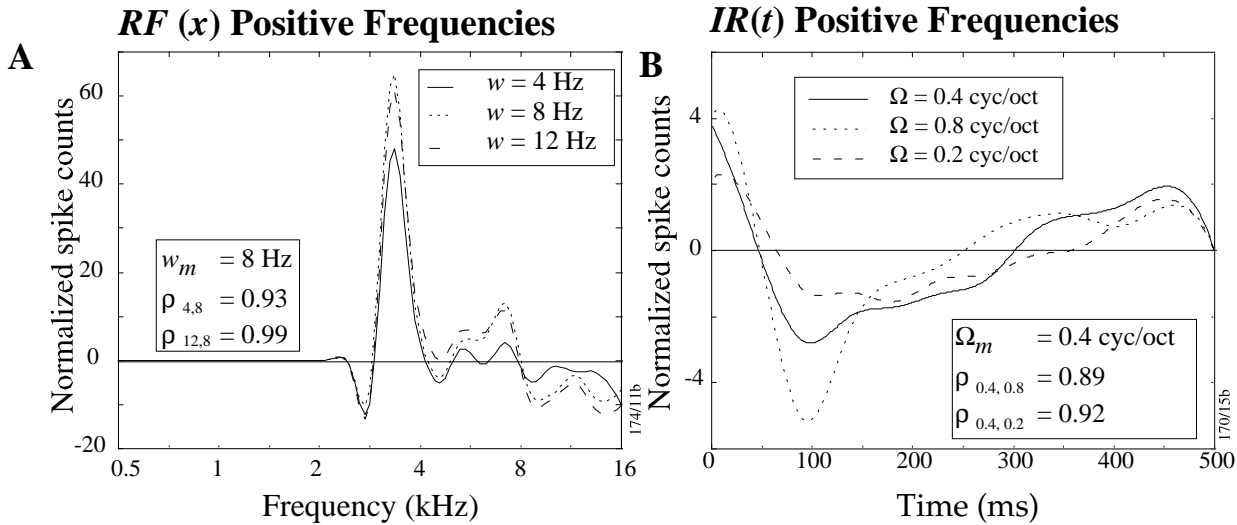}}
\caption[Experimental measurement of one-quadrant separability]
{\small\spacing{1} \sf
Left The positive-frequency RF computed at constant ripple velocity
for 3 different ripple velocities. The shapes should be the same if the
system is separable. Right The positive-frequency IR at constant ripple
frequency for 3 different ripple frequencies. The shapes should be the
same if the system is separable.}
\label{fig15}
\end{figure}

\subsection{Quadrant Linearity}

Linearity has been verified by presenting cells with a combination
of ripples from different quadrants${}^{16,22,23}$. As shown in
Figure~\ref{fig16} for one cell, the correlation between the predicted
and the measured response is (as in most cases) very good. Note that
the predicted response is shown in its non-half-wave rectified version:
as cells do not have negative firing rates, and the pentobarbital
anesthetic has reduced the spontaneous activity to zero, the comparison
should be made between the actual response and the half-wave rectified
version of the predicted response. The correlation coefficient $\rho$
in Figure~\ref{fig16} is the cross-correlation between the measured and
the predicted response. We have previously presented the correlation
between prediction and response within a single quadrant for 55 cells
and found 84\% of the cells with $\rho>0.6$.${}^{22}$ The error bars on
the measured response show the variability of cortical cells' responses
from sweep to sweep. Disparity is maximal between the prediction and the
actual spike count where both are small.

\begin{figure}
\centerline{\epsfysize=565pt
\epsfbox{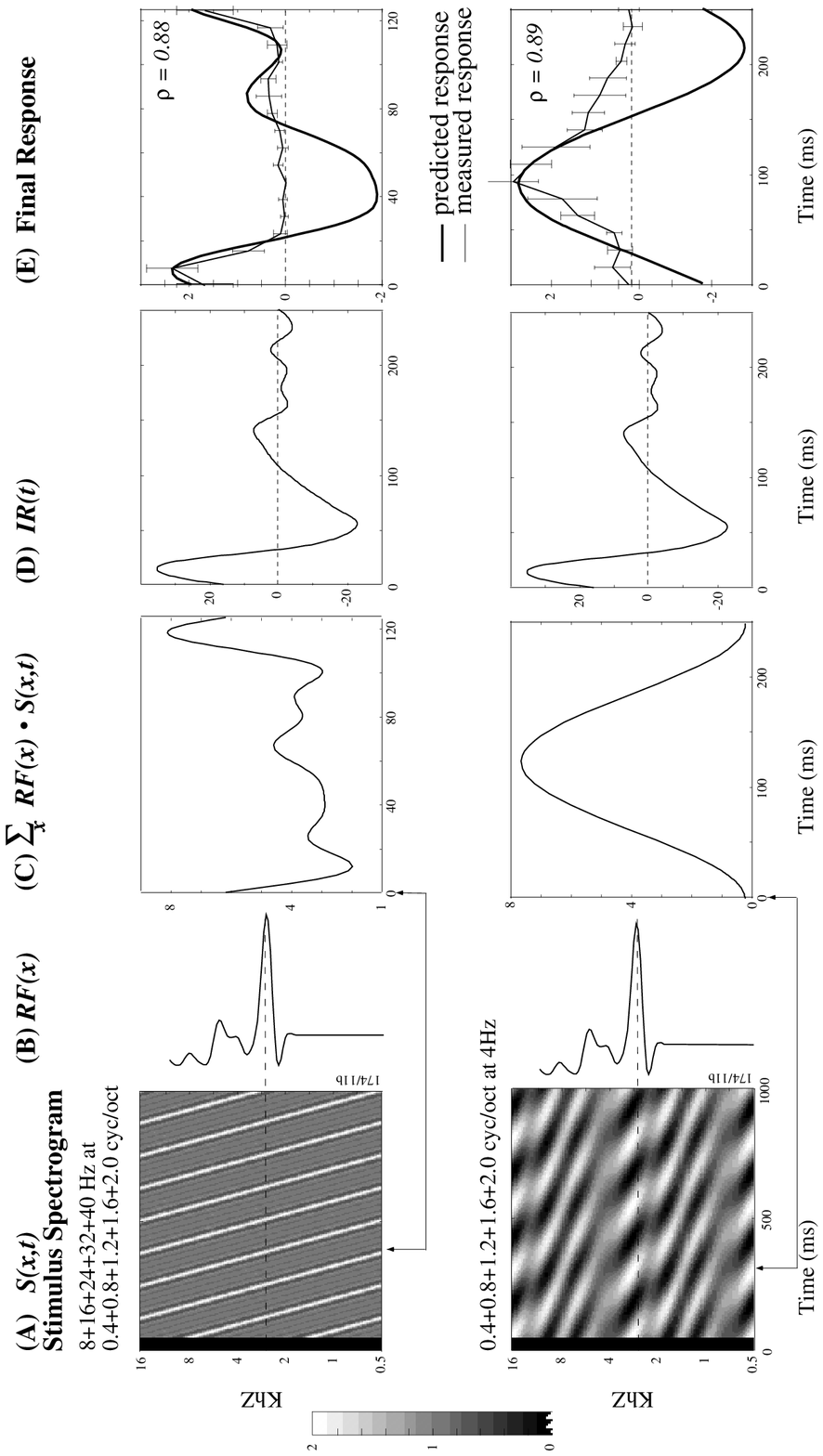}}
\caption[Experimental measurement of one-quadrant linearity]
{\small\spacing{.8} \sf
Linearity and separability within one quadrant. The stimulus (spectrogram
in A) is presented to a cell (RF shown in B). To predict the response, we
multiply the spectrogram with the RF (C), convolve the result with the
IR in D, obtaining the predicted response in E, the half-wave rectified
version of which should be compared with the actual response. 
Spontaneous activity is zero in this anesthetized preparation, and that
the ordinate axis is in arbitrary coordinates, except for the measured
response in E which is in spikes/sec.}

\label{fig16}
\end{figure}

\subsection{Full-Quadrant Separability and Linearity}

The remainder of this discussion describes logical extensions that
are currently under study. Thus far we have only verified separability
in a single quadrant. In vision, some cortical simple cells are fully
separable${}^{24}$, but all are at least quadrant separable${}^{25}$. We
have found both types in the auditory cortex as well; Figure~\ref{fig17}
shows examples of each. A fully separable cell has an STRF that is a
simple product of an RF and an IR, as in A. A quadrant separable cell, as
in B, does not, since it has different responses for upward and downward
moving ripples (as can be seen by inspection of its $STRF(x,t)$: it is
not symmetric about $x_m$). The separability of a cell does not affect
the linearity of responses to ripple combinations.

\begin{figure}
\centerline{
\epsfbox{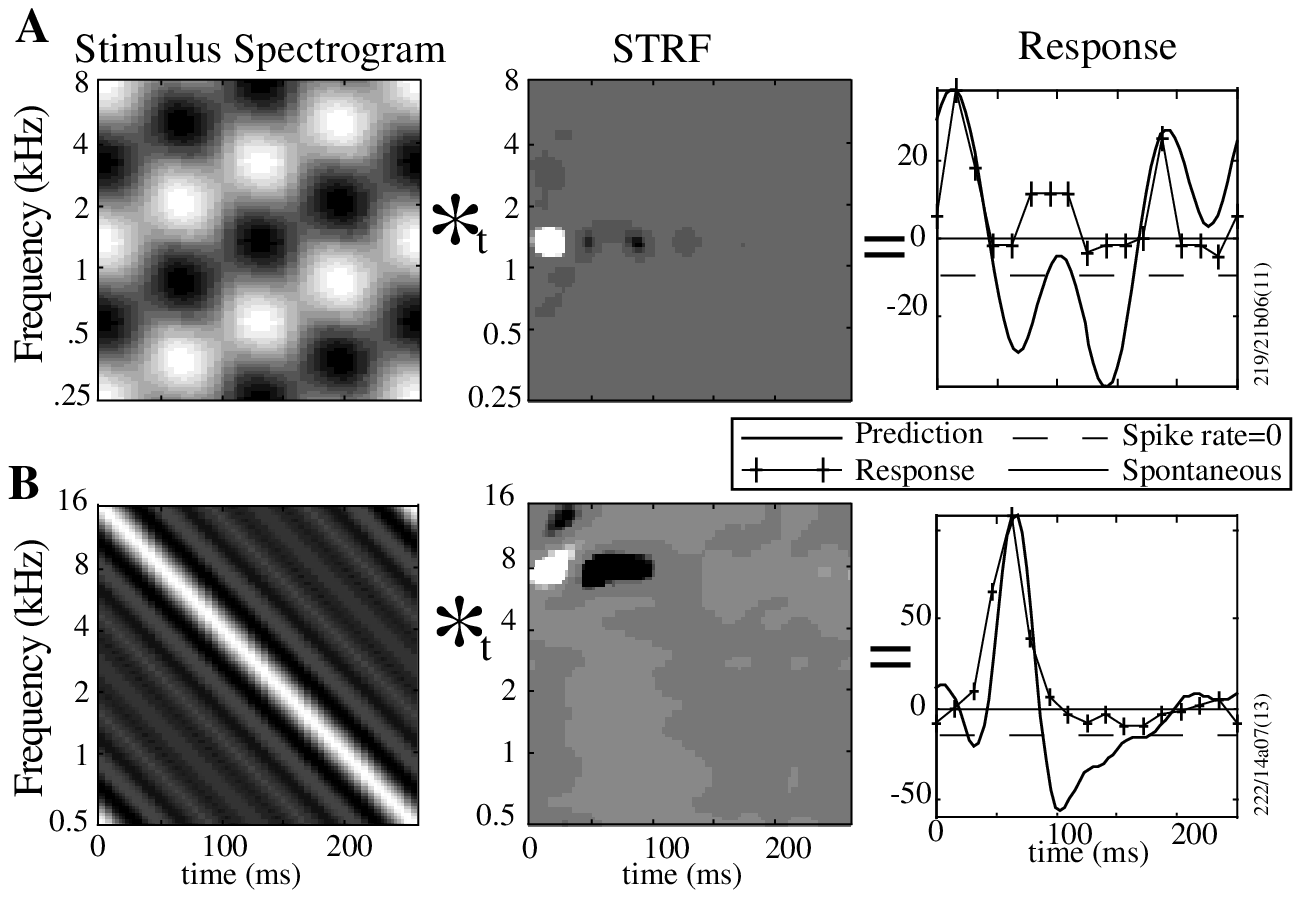}}
\caption[Experimental measurement of quadrant linearity]
{\small\spacing{1} \sf
Predictions of responses to complex dynamic spectra using the STRF. A The
predicted response is computed by a convolution (along the time dimension)
of the STRF with the spectrogram. The stimulus shown is composed of two
ripples (0.4 cycles/octave at 12 Hz and -4 Hz). The predicted waveform
is shown juxtaposed to the actual response (crosses) over one period of
the stimulus, in spikes/bin summed over 30 sweeps. B Another example: the
stimulus consists of a combination of ripples with ripple frequencies 0.2
cycles/octave at 4 Hz, 0.4 cycles/octave at 8 Hz, ... 1.2 cycles/octave
at 24 Hz, in cosine phase, resulting in an FM-like stimulus. In this
Ketamine/Xylazine preparation, the spontaneous activity was non-zero.}
\label{fig17}
\end{figure}

\subsection{Response Characteristics}

The transfer function for a specific cell is typically tuned to
a characteristic ripple frequency and velocity. The population of
cells shows a wide range of characteristic ripple frequencies and
velocities. Characteristic ripple velocities are mostly in the 8 - 16 Hz
range, rarely exceeding 30 Hz, and characteristic ripple frequencies are
mostly in the 0.4 - 0.8 cycles per octave range, rarely exceeding 2 cycles
per octave (in this anesthetized preparation). The slope of the transfer
function as a function of ripple frequency, $x_m$, corresponds to the center
frequency of the spectral envelope, which ranges from 200 Hz to at least
24 kHz (above which our acoustic delivery system is inadequate). The slope
of the transfer function as a function of ripple velocity, $\tau_d$, corresponds
to the center of the temporal envelope, which ranges roughly from 10
ms to 60 ms. The RF symmetry $\phi$, which describes the effects of lateral
inhibition and excitation, ranges roughly from $-90^o$ to $+90^o$ (out of
a possible $-180^o$ to $+180^o$), clustered around $0^o$. The IR polarity
$\theta$, which describes the polarity of the temporal response, ranges roughly
from $45^o$ to $135^o$ (out of a possible $0^o$ to $180^o$).

\section{Conclusions}

The emphasis in this review has been on presenting a technique to describe
neural response patterns of units in the cortex. More precisely, we use
moving ripples to characterize the response fields of auditory cortical
neurons, although this is a general method that can be used anywhere
responses are shown to be substantially linear for broadband stimuli.

Practically, we find that because of linearity of cortical responses
with respect to spectral envelope, we can use the ripple method to
characterize auditory cortical cell responses to dynamic, broadband
sounds. The linearity of the cortical unit responses is quantified by the
correlation coefficient between the predicted and the measured responses
curves. While at this point we do not have statistics to quantify the
linearity of response to ripples moving in both directions, linearity
within one quadrant (to down-moving ripples) has been extensively
quantified${}^{22}$, and we have no reason to expect linearity be any
different for ripples moving in both directions. The separability of
cells makes the ripple method practical, because of the time needed to
characterize a cell. One advantage of the method is the simultaneous
probing of spectral and temporal characteristics. Temporal processing
is becoming more and more recognized as an essential part of cortical
function, and the ripple method places it on an equal footing with spectral
processing. A caveat is that, thus far, the method only has been applied
to the steady state (i.e. periodic) response of cells.

We find that response fields in AI tend to have characteristic shapes both
spectrally and temporally. Specifically, AI cells are tuned to moving
ripples, i.e., a cell responds well only to a small set of moving ripples
around a particular spectral peak spacing and velocity. We find cortical
cells with all center frequencies, all spectral symmetries, bandwidths,
latencies and temporal impulse response symmetries. One way to interpret
this result is that AI decomposes the input spectrum into different
spectrally and temporally tuned channels. Another equivalent view is
that a population of such cells, tuned around different moving ripple
parameters, can effectively represent the input spectrum at multiple
scales. For example, spectrally narrow cells will represent the fine
features of the spectral profile, whereas broadly tuned cells represent
the coarse outlines of the spectrum. Similarly, dynamically sluggish cells
will respond to the slow changes in the spectrum, whereas fast cells
respond to rapid onsets and transitions. In this manner, AI is able to
encode multiple different views of the same dynamic spectrum. From this,
we conclude that the primary auditory cortex performs multi-dimensional,
multi-scale wavelet transform of the auditory spectrum.

Pitch is very important to the auditory system. The spectral ripple
responses presented here do not have pitch, since they are synthetized
with logarithmically spaced carrier tones. We have not yet examined
unit responses to a ripple spectra with harmonically related carrier
tones. Consequently, all our unit responses are due to the envelope or
spectral profile of the broadband stimulus, and are not dependent on the
carrier tones. It is quite possible that the pitch of a harmonic series
of tones will affect the responses. It is also possible that sufficiently
narrowly tuned cells might directly encode the harmonic spacing in a
spectrum in a systematic manner to encode the pitch as was discussed in
detail in Wang and Shamma${}^{26}$. This is work in progress.

The suggestion that cortical cells are linear might appear far-fetched
given the non-linear response to pure tones, such as rate vs. intensity
functions with threshold, saturation, and non- monotonic behavior (Brugge
and Merzenich${}^{27}$; Nelken et al.${}^{8}$). Nevertheless, we find that
the non-linearity observed with broadband ripple spectra is substantially
smaller than with tonal stimuli, when it comes to predicting the response
of a cell to a combination of stimuli, knowing the response to individual
ones. Furthermore, just as measuring linear systems response properties
with tones, such as bandwidth, rate-level functions, tuning quality
factor and other measures is considered meaningful, characteristics of
the ripple responses prove useful, and relate to the properties measured
with tones${}^{18,16}$. Investigations currently under way in the Inferior
Colliculus will shed light on the mechanisms that allow cells to exhibit
a linear behavior in auditory cortex, so many synapses away from the
auditory nerve.

\section{Acknowledgements}

This work is supported by a MURI grant N00014-97-1-0501 from the Office
of Naval Research, a training grant NIDCD T32 DC00046-01 from the National
Institute on Deafness and Other Communication Disorders, and a grant NSFD
CD8803012 from the National Science Foundation. We would like to thank
David Klein, Izumi Ohzawa and Alan Saul.

\section{References}

\goodbreak \frenchspacing \parskip=0pt \renewcommand{\baselinestretch}{1}\small
\par

\indent\Item{1.} M.M. Merzenich, P.L. Knight and G.L. Roth, Representation
of the cochlear partition on the superior temporal plane of the macaque
monkey, Brain Res. 50, 231-249 (1975). R.A. Reale and T.J. Imig, Tonotopic
organization in auditory cortex of the cat, J. Comp. Neurol. 192, 265-291
(1980).
\Item{2.} J.C. Middlebrooks, R.W. Dykes and M.M. Merzenich, Binaural
response-specific bands in primary auditory cortex of the cat:
topographical organization orthogonal to isofrequency contours, Brain
Res. 181, 31-48 (1980).
\Item{3.} S.A. Shamma, J.W. Fleshman, P.R. Wiser and H. Versnel,
Organization of response areas in ferret primary auditory cortex,
J. Neurophys. 69, 367-383 (1993).
\Item{4.} C.E. Schreiner, J. Mendelson and M.L. Sutter, Functional
topography of cat primary auditory cortex: representation of tone
intensity, Exp. Brain Res. 92, 105-122 (1992).
\Item{5.} C.E. Schreiner and M.L. Sutter, Topography of excitatory
bandwidth in cat primary auditory cortex: single- versus multiple-neuron
recordings, J. Neurophysiol. 68, 1487-1502 (1992).
\Item{6.} J. Mendelson, C.E. Schreiner, M.L. Sutter and K. Grasse,
Functional topography of cat primary auditory cortex: selectivity to
frequency sweeps, Exp. Brain Res. 94, 65-87 (1993).
\Item{7.} R.L. De Valois and K.K. De Valois, Spatial Vision, Oxford
University Press, New-York (1988).
\Item{8.} I. Nelken, Y. Prut, E. Vaadia and M. Abeles, Population
responses to multifrequency sounds in the cat auditory cortex: One- and
two-parameter families of sounds, Hear. Res. 72, 206-222 (1994).
\Item{9.} M.L. Sutter,.W.C. Loftus and K.N. O'Connor, Temporal
properties of two-tone inhibition in cat primary auditory cortex, ARO
midwinter meeting (1996).
\Item{10.} L.R. Rabiner and R.W. Schafer, Digital processing of
speech signals, Prentice-Hall, New-Jersey (1978).
\Item{11.} J.E. Rose, J.F. Brugge, D.J. Anderson and J.E. Hind,
Phase-locked response to low-frequency tones in single auditory nerve
fibers of the squirrel monkey, J. Neurophys. 30, 769-793 (1967).
\Item{12.} W.S. Rhode and P.H. Smith, Encoding time and intensity
in the central cochlear nucleus of the cat, J. Neurophys. 56, 262-286
(1986).
\Item{13.} W.S. Rhode and S. Greenberg, Physiology of the cochlear
nuclei, in The mammalian auditory pathway: Neurophysiology, Popper,
A.N. and Fay, R.R. editors, Springer-Verlag, New-York (1991).
\Item{14.} R. Lyon and S.A. Shamma, Timbre and pitch, in Auditory
computation, H.L. Hawkins., T.A. McMullen, A.N. Popper and R.R. Fay,
editors, Springer-Verlag, New-York (1995).
\Item{15.} G. Langner, Periodicity coding in the auditory system,
Hearing Res. 6, 115-142 (1992). D.A. Depireux, D.J. Klein, J.Z. Simon
and S.A. Shamma, Neuronal correlates of pitch in the Inferior Colliculus,
ARO midwinter meeting (1997).
\Item{16.} N. Kowalski, D.A. Depireux and S.A. Shamma, Analysis of
dynamic spectra in ferret primary auditory cortex: I. Characteristics
of single unit responses to moving ripple spectra, J. Neurophys. 76,
(5) 3503-3523 (1996).
\Item{17.} D.K. Ryugo, The auditory nerve: peripheral innervation,
cell body morphology, and central projections, in The mammalian auditory
pathway: neuroanatomy, D.B. Webster, A.N. Popper, and R.R. Fay, editors,
Springer-Verlag, New-York, (1991).
\Item{18.} C.E. Schreiner and B.M. Calhoun, Spatial frequency filters in
cat auditory cortex. Auditory Neurosci. 1, 39-61 (1994). S.A. Shamma,
H. Versnel and N. Kowalski, Ripple analysis in ferret primary auditory
cortex: I. Response characteristics of single units to sinusoidally
rippled spectra. Auditory Neurosci. 1, 233-254 (1995). S.A. Shamma
and H. Versnel, Ripple analysis in ferret primary auditory cortex:
II. Prediction of unit responses to arbitrary spectral profiles. Auditory
Neurosci. 1, 255-270 (1995). H. Versnel, N. Kowalski and S.A. Shamma,
Ripple analysis in ferret primary auditory cortex: III. Topographic
distribution of ripple response parameters. Auditory Neurosci. 1, 271-285
(1995).
\Item{19.} A. Papoulis, The Fourier integral and its applications, McGraw-Hill
(1962).
\Item{20.} L. Cohen, Time-frequency analysis, Prentice-Hall, New Jersey
(1995).
\Item{21.} D. Dong and J.J. Atick, Temporal decorrelation: a theory of
lagged and nonlagged cells in the lateral geniculate nucleus, Network:
Computation in Neural Systems 6, 159-178 (1995).
\Item{22.} N. Kowalski, D.A. Depireux and S.A. Shamma, Analysis of dynamic
spectra in ferret primary auditory cortex: II. Prediction of unit responses
to arbitrary dynamic spectra. J. Neurophys. 76, (5) 3524--3534 (1996).
\Item{23.} J.Z. Simon, D.A. Depireux and S.A. Shamma, Representation
of complex dynamic spectra in auditory cortex, in Psychophysical and
physiological advances in hearing, A.R. Palmer, A.Rees, A.Q. Summerfield
and R. Meddis, editors, Whurr Publishers, London (1988).
\Item{24.} J. McLean and L.A. Palmer, Organization of simple cell responses
in the three-dimensional frequency domain. Vis. Neurosc. 11, 295-306
(1994). G.C. DeAngelis, I. Ohzawa and R.D. Freeman, Receptive-field
dynamics in the central visual pathways. Trends Neurosc. 18, 451-458
(1995).
\Item{25.} B.W. Andrews and D.A. Pollen, Relationship between spatial
frequency selectivity and receptive field profile of simple cells,
J. Physiol. (London) 287, 163-176 (1979). S.M. Friend and C.L. Baker,
Spatio-temporal frequency separability in area 18 neurons of the cat,
Vision Res. 33, 1765-1771 (1993).
\Item{26.} K. Wang and S. A. Shamma, Self-normalization and noise robustness
in auditory representations. IEEE Trans. Audio. Speech. Proc. 2, 421-435
(1994).
\Item{27.} J. F. Brugge and M. M. Merzenich, Responses of neurons
in auditory cortex of the Macaque monkey to monaural and binaural
stimulation. J. Neurophysiol. 36, 1138-1158 (1973).

\end{document}